\documentclass[twocolumn]{aastex63}

\newcommand{\gppr}{\stackrel{>}{\scriptstyle \sim}}
\newcommand{\gappr}{\raisebox{-0.4ex}{$\gppr$}}
\received{May, 2020}
\revised{, 2020}
\accepted{\today}
\submitjournal{ApJL}

\shorttitle{How Jupiters change the fate of Neptunes}
\shortauthors{Ronco et al.}
\graphicspath{{./}{figures/}}
\begin{document}

\title{How Jupiters save or destroy inner Neptunes around evolved stars}

\correspondingauthor{M.\,P. Ronco, M.\,R. Schreiber}
\email{mronco@astro.puc.cl, matthias.schreiber@uv.cl}

\author[0000-0003-1385-0373]{Mar\'{\i}a Paula Ronco}
\affil{Instituto de Astrof\'{\i}sica - Pontificia Universidad Cat\'olica de Chile,
Av. Vicu\~na Mackenna 4860, Macul - Santiago, 8970117 , Chile.}
\affil{Millennium Nucleus for Planet Formation, NPF, Chile.}

\author[0000-0003-3903-8009]{Matthias R. Schreiber}
\affil{Instituto de F{\'i}sica y Astronom{\'i}a, Universidad de Valpara{\'i}so, Valpara{\'i}so, Chile.}
\affil{Millennium Nucleus for Planet Formation, NPF, Chile.}
%\affiliation{}

%\collaboration{1}{(AAS Journals Data Scientists collaboration)}

\author[0000-0002-7460-3264]{Cristian A. Giuppone}
\affil{Universidad Nacional de C\'ordoba, Observatorio Astron\'omico - IATE. Laprida 854, 5000 C\'ordoba, Argentina.}
%\affiliation{}
%\nocollaboration{1}

\author[0000-0001-8014-6162]{Dimitri Veras}
\affil{Centre for Exoplanets and Habitability, University of Warwick, Coventry CV4 7AL, UK.}
\affil{Department of Physics, University of Warwick, Coventry CV4 7AL, UK.}

%\collaboration{1}{(LaTeX collaboration)}

\author[0000-0003-1965-3346]{Jorge Cuadra}
\affil{Departamento de Ciencias, Facultad de Artes Liberales, Universidad Adolfo Ib\'a\~nez, Avenida Padre Hurtado 750, Vi\~na del Mar, Chile.}
\affil{Millennium Nucleus for Planet Formation, NPF, Chile.}

\author[0000-0001-8577-9532]{Octavio M. Guilera}
\affil{Instituto de Astrof\'{\i}sica de La Plata, CONICET-UNLP, La Plata, Argentina.}
\affil{Instituto de Astrof\'{\i}sica - Pontificia Universidad Cat\'olica de Chile,
Av. Vicu\~na Mackenna 4860, Macul - Santiago, 8970117 , Chile.}
\affil{Millennium Nucleus for Planet Formation, NPF, Chile.}
\nocollaboration{6}

%% Note that the \and command from previous versions of AASTeX is now
%% depreciated in this version as it is no longer necessary. AASTeX 
%% automatically takes care of all commas and "and"s between authors names.

%% AASTeX 6.3 has the new \collaboration and \nocollaboration commands to
%% provide the collaboration status of a group of authors. These commands 
%% can be used either before or after the list of corresponding authors. The
%% argument for \collaboration is the collaboration identifier. Authors are
%% encouraged to surround collaboration identifiers with ()s. The 
%% \nocollaboration command takes no argument and exists to indicate that
%% the nearby authors are not part of surrounding collaborations.

%% Mark off the abstract in the ``abstract'' environment. 
\begin{abstract}
In about 6 Giga years our Sun will evolve into a red giant and finally end its life as a white dwarf. This stellar metamorphosis will occur to virtually all known host stars of exo-planetary systems and is therefore crucial for their final fate. It is clear that the innermost planets will be engulfed and evaporated during the giant phase and that planets located farther out will survive. However, the destiny of planets in-between, at $\sim\,1-10$\,au, has not yet been investigated with a multi-planet tidal treatment. We here combine for the first time multi-planet interactions, stellar evolution, and tidal effects in an $N$-body code to study the evolution of a Neptune-Jupiter planetary system. We report that the fate of the Neptune-mass planet, located closer to the star than the Jupiter-mass planet, can be very different from the fate of a single Neptune. The simultaneous effects of gravitational interactions, mass loss and tides can drive the planetary system towards mean motion resonances. Crossing these resonances affects particularly the eccentricity of the Neptune and thereby also its fate, which can be engulfment, collision with the Jupiter-mass planet, ejection from the system, or survival at a larger separation.
\end{abstract}

%% Keywords should appear after the \end{abstract} command. 
%% See the online documentation for the full list of available subject
%% keywords and the rules for their use.

%% From the front matter, we move on to the body of the paper.
%% Sections are demarcated by \section and \subsection, respectively.
%% Observe the use of the LaTeX \label
%% command after the \subsection to give a symbolic KEY to the
%% subsection for cross-referencing in a \ref command.
%% You can use LaTeX's \ref and \label commands to keep track of
%% cross-references to sections, equations, tables, and figures.
%% That way, if you change the order of any elements, LaTeX will
%% automatically renumber them.
%%
%% We recommend that authors also use the natbib \citep
%% and \citet commands to identify citations.  The citations are
%% tied to the reference list via symbolic KEYs. The KEY corresponds
%% to the KEY in the \bibitem in the reference list below. 

\section{Introduction} \label{sec:intro}

More than 4000 exoplanets have been confirmed so far\footnote{ \citep[\url{http://exoplanet.eu/}, ][]{Schneider2011}}
  and the discovered planetary systems, the vast majority around sun-like main sequence stars, reveal a great variety in terms
  of the number of planets, their masses, and  orbital separations. These planetary systems are often referred to as the
  \emph{final} outcome of planet formation, and are typically compared to the predictions of population synthesis analysis
  \citep{Ronco2017,Mordasini2018} and/or $N$-body simulations \citep[e.g.][]{Pfyffer2015, Ronco2018}.

  However, the evolution of stars does not end on the main sequence and therefore neither does the evolution of the
  planetary systems around them. More than 100 gas giant
  planets\footnote{\url{https://www.lsw.uni-heidelberg.de/users/sreffert/giantplanets/giantplanets.php}} have been
  discovered around red giant stars \citep[e.g.][]{Jones2016} and convincing evidence for the existence of planetary debris
  and planets around white dwarfs has been provided in the last decades. About one third of all white dwarfs show atmospheric
  metal absorption lines that must result from the recent accretion of solid material \citep{Koester2014}. Roughly $5$ per
  cent of these white dwarfs show a detectable infrared excess indicative of the presence of a circumstellar debris disk
  \citep{Barber2012}, and about the same fraction of the latter show a detectable gaseous disk component \citep{Manser2020}.
  As first suggested by \citet{Jura2003}, metal polluted white dwarfs and the disks around them are the result of the tidal
  disruption of rocky planetary material \citep{Veras2014, Malamud2020}. In the spectacular case of the transiting and
  disintegrating planetesimal around WD\,1145+017 we can witness this process in real time \citep{Vandenburg2015,Gaensicke2016}.
  Additional recent discoveries related to the final fate of planetary systems include a planetesimal which might resemble
  the core of an disrupted Earth-like planet orbiting a white dwarf in a close orbit \citep{Manser2019} and a close-in
  Neptune-like planet that is evaporated by EUV irradiation from the white dwarf \citep{Gansicke2019} and was possibly tidally
  disrupted during its orbital decay \citep{Veras2020}. For hot white dwarfs, \citet{Schreiber2019} showed that the observed
  metal pollution \citep{Barstow2014} can be explained if a large fraction ($\sim50$ per cent) of white dwarfs host planets
  at separations 
  $\gappr3$\,au. This prediction is in line with the results of microlensing planet surveys that predict a
  large number of planets, mostly Neptunes, beyond the snow line \citep{Suzuki2016}.

  Understanding the formation of planetary debris and the existence of planets around white dwarfs requires modelling the
  evolution of planetary systems beyond the main sequence \citep[e.g.][]{Veras2015b}. When the host star evolves into a giant
  star, each planet in a surrounding planetary system is subject to orbital changes. These changes are dominated by stellar
  tides and stellar mass-loss, particularly for planets inside $\sim$\,10~au, although other mechanisms can have additional
  minor effects on the planetary orbits \citep[see][and references therein]{Veras-Review2016}. Stellar mass-loss tends to
  expand the orbit of the planets due to a decrease of the gravitational potential and simultaneous conservation of angular
  momentum but leaves the eccentricities nearly unchanged. Stellar tides, in contrast, tend to decrease both the semi-major axis
  and eccentricities of planetary orbits. 

  Several studies have determined the conditions under which {\em{single}} planets can survive the evolution of their
  host star into a white dwarf. However, the number of discovered planetary systems with more than just one planet is
  continuously increasing \citep[e.g.][]{Shallue2018} and it seems plausible to assume that such systems might be the rule rather than the exception. 
  It is then crucial to simulate the evolution of multiple planets affected by mass loss and
  stellar tides taking into account their mutual gravitational interactions. This has not been done yet. 

  We here close this gap by analysing the dynamical evolution of hypothetical planetary systems consisting of an inner
  Neptune- and an outer Jupiter-mass planet when their central star evolves through the Red Giant Branch (RGB) taking into
  account stellar mass-loss, stellar tides and the mutual gravitational interactions between the two planets. We find that
  under certain conditions the fate of the Neptune-mass planet is indeed significantly affected by the presence of the
  outer Jupiter-mass planet. Most interestingly, a Neptune-mass planet that would survive without a companion can be pushed into
  the giant star and a Neptune-mass planet that would not survive alone can be saved by the outer Jupiter.
  
\section{Physical Model} \label{sec:Model}

In order to calculate the rate of change of the semi-major axis of a planet affected by the stellar mass-loss and stellar
tides we adopt the formalism by \citet{Zahn1977}, which was also used by \citet{Villaver2009} and \citet{Villaver2014}.
According to these authors, the change in orbital separation can be written as
\begin{eqnarray}
  \left( \frac{\dot{a}}{a} \right) = -\left( \frac{\dot{M_\star}}{M_\star+M_{\text{p}}} \right)-\left(\frac{\dot{a}}{a}\right)_{\text{t}},
\label{eq:evol-semieje}
\end{eqnarray}

where $M_\star$ is the total mass of the star, $M_{\text{p}}$ the mass of the planet, $a$ the semi-major axis of the planet's
orbit, and $(\dot{a}/{a})_{\text{t}}$ denotes the change of the semi-major axis caused by stellar tides given by:
\begin{eqnarray}
\label{eq:semieje-tides}
  {\left(\frac{\dot{a}}{a}\right)_{\text{t}}} = -\frac{1}{9\tau_{\text{conv}}}\frac{M^{\text{env}}_\star}{M_\star}\frac{M_{\text{p}}}{M_\star}\left(1+\frac{M_{\text{p}}}{M_\star}\right)\left(\frac{R_\star}{a}\right)^8 \\ \nonumber
  \times\left[2p_2 + e^2\left(\frac{7}{8}p_1-10p_2+\frac{441}{8}p_3\right)\right].
\end{eqnarray}
Here $M^{\text{env}}_\star$ is the envelope mass and $R_\star$ the radius of the star. We follow \citet{Rasio1996} to compute
$\tau_{\text{conv}}$, which is the eddy turnover timescale within the stellar envelope which writes as:
\begin{eqnarray}
 \tau_{\text{conv}} = \left[\frac{M^{\text{env}}_\star R^{\text{env}}_\star (R_\star-R^{\text{env}}_\star)}{3L_\star}\right]^{1/3},
 \label{eq:turnover_timescale}
\end{eqnarray}
where $R^{\text{env}}_\star$ is the radius at the base of the convective envelope and $L_\star$ the luminosity of the star.
The frequency components of the tidal force, i.e. $p_1$, $p_2$ and $p_3$ are, as in \citet{MustillVillaver2012}, given by:
\begin{eqnarray}
 p_i \approx \frac{9}{2} {\text{min}}\left[1, \left( \frac{4\pi^2 a^3}{i^2G(M_\star+M_{\text{p}})\tau^2_{\text{conv}}} \right)\right], i = 1,2,3.
 \label{eq:frequencies}
\end{eqnarray}
In a similar way, the eccentricity rate of change for the planet generated by tidal forces can be written as
\begin{eqnarray}
\label{eq:excentricidad-tides}
  \left(\frac{\dot{e}}{e}\right)_{\text{t}} = -\frac{1}{36\tau_{\text{conv}}}\frac{M^{\text{env}}_\star}{M_\star}\frac{M_{\text{p}}}{M_\star}\left(1+\frac{M_{\text{p}}}{M_\star}\right)\left(\frac{R_\star}{a}\right)^8 \\ \nonumber
  \times\left[\frac{5}{4}p_1 -2p_2 +\frac{147}{4}p_3\right].
\end{eqnarray}
  Other mechanisms such as possible changes in the planet's mass due to evaporation of its surface by EUV radiation or due
  to the accretion of a fraction of the ejected stellar material \citep{Villaver2009,Villaver2014}, drag forces that occur
  when the planet passes through the gas expelled from the host star, and planetary tides, have been taken into account previously. 
  However, these additional forces and the corresponding changes in the planet mass and orbital parameters are negligible
  compared to the effects of stellar mass-loss and stellar tides \citep{Villaver2009, VerasEgglGansicke2015,Rao2018}. 
  We therefore only consider the latter. 

\section{Numerical Methods} \label{sec:Methods}

In what follows we describe the numerical tools we used to calculate the evolution and fate of a Neptune-mass planet
affected by the stellar mass-loss, stellar tides, and mutual gravitational interactions with a Jupiter-mass planet during the RGB.
  
\subsection{The stellar evolution code}

We use the stellar evolution code {\tt SSE} developed by \citet{Hurley2000} which produces a single evolutionary track from
a set of zero-age-main-sequence values such as the mass of the star $M_\star$, the metallicity $Z$, and the Reimers parameter
  $\eta$ which controls the mass-loss rate \citep{Reimers1975}. As we are only interested in the main effect of combining tidal
  forces, mass loss, and gravitational interactions we fixed the stellar evolution parameters to $M_\star=1M_\odot$, $Z=0.02$,
  and $\eta = 0.5$. The resulting evolutionary track provides information about the main parameters of the host star such as
  $M_\star$, $R_\star$, $M^{\text{env}}_\star$, $R^{\text{env}}_\star$ and $L_\star$ as a function of time from the zero-age-main-sequence
  and until the star becomes a white dwarf. Here we only discuss planetary system dynamics before and just beyond the tip of the RGB, during which the maximum radius achieved by the star is $\sim 186.34R_\odot$, equivalent to $\sim$ 0.86 au, and the total stellar mass loss accumulates to $\sim 0.24M_\odot$.
  Analyses that include the AGB and WD phases will be presented in future papers. 
\subsection{Evolution of a single planet}
\label{sec:RKF-integrator}

\begin{figure}
  \centering
  \includegraphics[width= 0.45\textwidth]{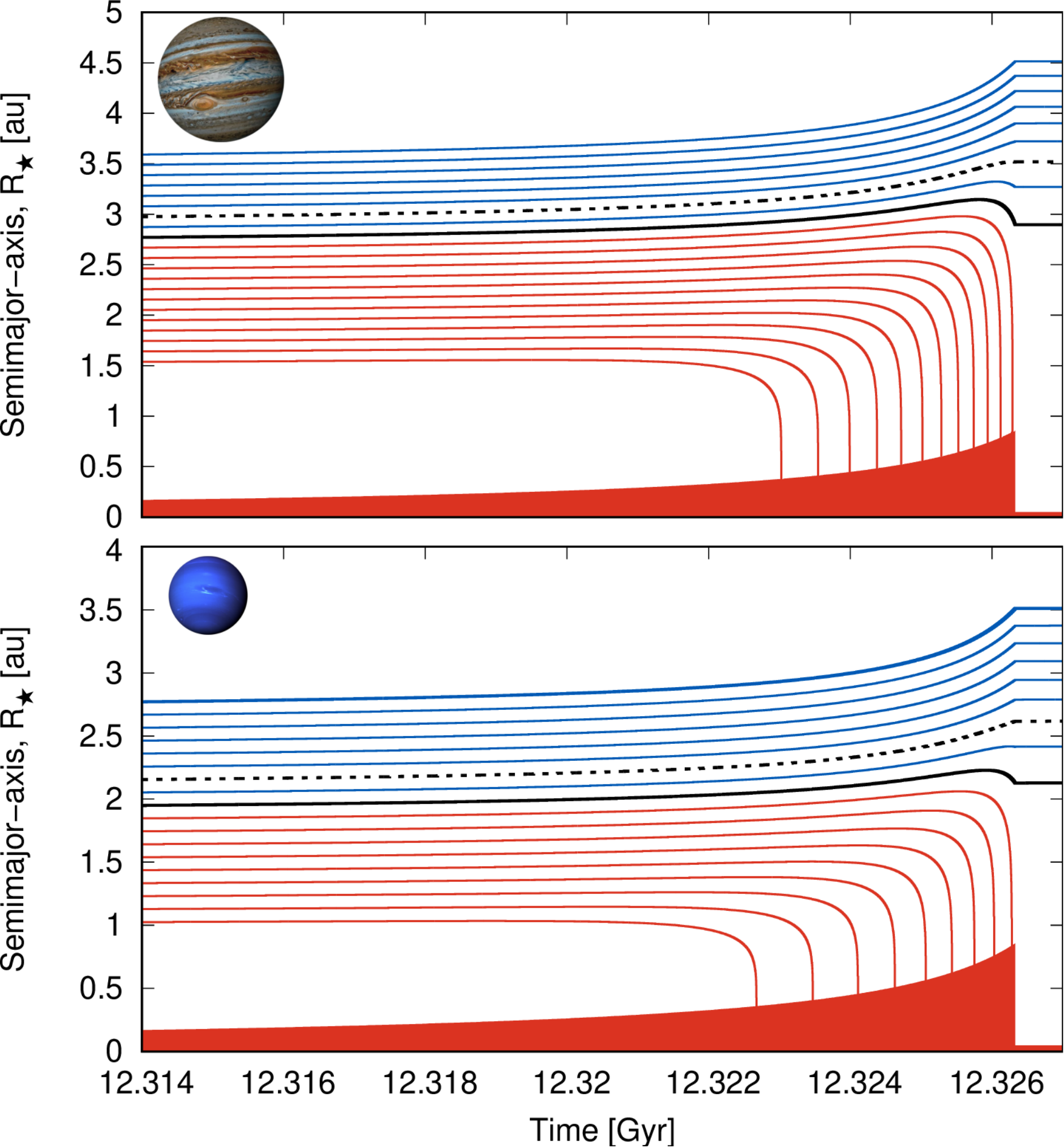}
  \caption{ Evolution of the semi-major axis for a Jupiter (top) and a Neptune-mass planet (bottom) assuming initial separations of 
    $1.5-3.5$~au and $1-2.7$~au, respectively with a step size of 0.1~au. Both planets are assumed to be in circular and coplanar
    orbits. The red filled area represents the radius of the host star for $\sim$ 0.013 Gyr of its evolution towards the end of
    the RGB. The red solid lines show evolutionary tracks that result in an engulfment while the blue ones show those that lead to
    survival of the planets. The black solid line represents the limit between engulfment and survival (at 2.7~au for the Jupiter
    and at 1.9~au for the Neptune-mass planet) while the dashed line provides the limit above which
    tides are not strong enough to generate orbital decay at any time.  
    %are mainly affected
    %by stellar mass-loss as stellar tides become negligible.
  }
  \label{fig:evolution-tracks}
\end{figure}

  We calculated the evolution of a single planet using a Runge-Kutta Fehlberg (RKF) algorithm to integrate
  Eq.\ref{eq:evol-semieje} coupled with Eq. \ref{eq:excentricidad-tides} while taking into account the previously determined
  stellar evolution track. This numerical tool allows us to rapidly compute the fate of a single planet after the RGB for
  a range of initial orbital and planetary parameters. As an example, Fig.\ref{fig:evolution-tracks} shows
  the time evolution of the semi-major axis of a Jupiter (top) and of a Neptune-mass planet (bottom), located at
  different initial positions in circular and coplanar orbits. 

We integrated the time evolution for initial separations between 1.5~au and 3.5~au for the Jupiter (top panel) and 1~au
  and 2.7~au for the Neptune-mass planet (bottom panel). The Jupiter-mass planets with an initial semi-major axis
  above 2.7~au survive while single Neptune-mass planets survive for initial separations exceeding 1.9~au.

  Although our model is simpler than previously performed simulations, our result for the survival of a Jupiter-mass planet with
  initial semi-major axis larger than 2.6\,au is in agreement with more detailed numerical calculations. Using a different
  expression for the convective timescale, a different evolutionary track for the central star, and considering drag forces
  (which we ignored), \citet{Villaver2009} found that a single Jupiter-mass planet around an evolving solar-mass star is engulfed
  for initial separations $a < 3~$au. According to \citet[][their figure\,3]{NordhausSpiegel2013}, who in addition to the most
  important forces considered changes in the primary spin, a Jupiter-mass planet around a solar-mass star would survive for initial
  separations exceeding $\sim2.5-3$~au.  

\subsection{$N$-body integrator}

  To model the dynamics of a planetary system affected by stellar evolution we use the modified version of the {\scriptsize{MERCURY}}
  integration package \citep{Chambers1999} developed by \citet{Veras2013b}, which uses the Bulirsch Stoer (BS) integrator. This code
  interpolates the {\tt SSE} code \citep{Hurley2000} stellar mass output at each {\scriptsize{MERCURY}} time-step and at each BS substep,
  which produces the same single evolutionary track for a Solar-type star as the one used with the RKF integrator described in
  Sec.\,\ref{sec:RKF-integrator}. For details about the computation of the mass-loss we refer the reader to \citet{Veras2013b}. 

  We implemented stellar tides in this version of {\scriptsize{MERCURY}} following the formalism described in
  Sec.\,\ref{sec:Model} as an external force so that planetary evolution is affected not only by gravitational interactions between
  planets and with the central star, but also by dissipative effects. The description of the implementation of the stellar tides and
  its validation with the RKF integrations can be found in the Appendix.
  
    \section{Results and Discussion}
\label{sec:Results}

  We analyze the evolution of a two-planet system formed by an inner Neptune and an outer Jupiter-mass planet. For simplicity, both
  planets are initially in circular and coplanar orbits. We calculated a grid with initial locations ranging from 2 to 3~au with a
  step size in separation of 0.1~au for the Jupiter and from 1.5 to 2.1~au with a step size of 0.05~au for the Neptune. This grid
  covers all combinations of fates obtained with the RKF integrations for single planets (engulfment--engulfment, engulfment--survival,
  survival--engulfment and survival--survival). For some of these semi-major axis pairs simulations are not performed because they
  violate the classical $\sim$ $3.5R_{\text{mH}}$ stability criteria \citep{Gladman1993}, where $R_{\text{mH}}$ is the mutual Hill radius \citep[see also][]{Giuppone+2013}.
  For the rest of the angular orbital parameters we adopt random values between $0^{\circ}$ and $360^{\circ}$. We integrate each configuration with both planets together for $\sim$ 750 Myr %(or 0.75 Gyr)
  starting at the base of the giant branch which according to {\tt SSE} is reached by a star with one solar mass at an age of $\sim$11.6 Gyr. We used an accuracy parameter of $1\times 10^{-13}$ \citep{Veras2013b} and saved the results every 1000
  years. Collisions are treated as inelastic mergers, and close encounters are defined and recorded within 3 Hill radii. 

\begin{figure}
  \centering
    \includegraphics[width= 0.48\textwidth]{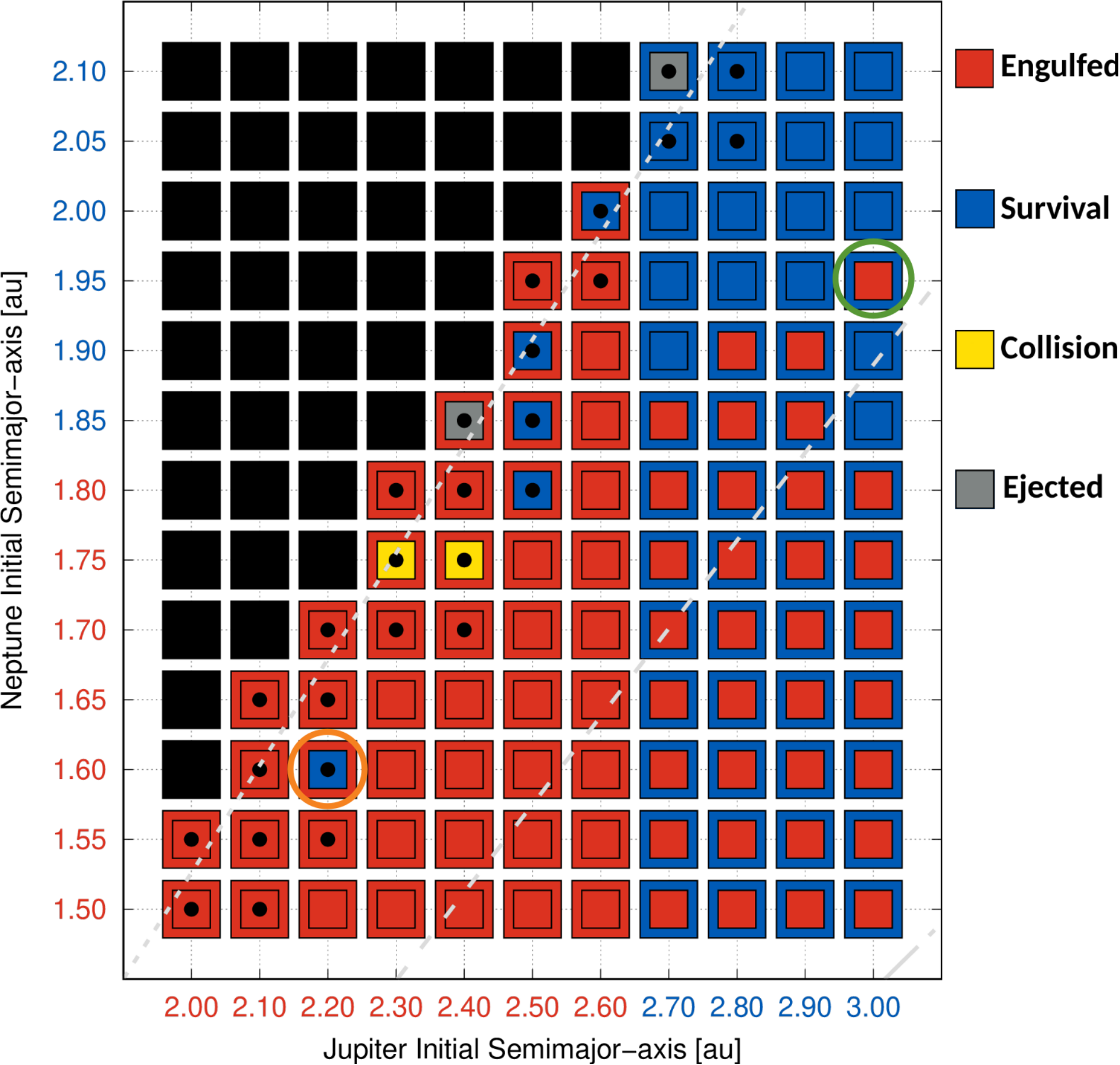}
    \caption{Final fate of the Jupiter (large squares) and the Neptune-mass planet (small inner squares) as a function of their
      initial separations. Red squares indicate the planet is engulfed by the central star, grey denotes the planet is ejected,
      blue means the planet survives, and yellow means the planet collides with the larger one. The black dots in the
      centre of some of the squares highlight those simulations in which close encounters (defined within 3 Hill radii)
      %a separation between the two plants smaller than 3 Hill radii of each planet
      took place, and the grey long and short dashed lines show the positions of the 2:1 and the 3:2 MMR, respectively. 
      The numbers providing the initial semi-major axis of both planets are coloured in red or blue, depending on whether they would
      survive or be engulfed as single planets. The filled black squares represent combinations that are unstable. Two of the
      cases where the presence of the outer planet changes the fate of the Neptune-mass planet are highlighted with green and orange circles. 
  }
  \label{fig:Resultados-Generales}
\end{figure}

Figure\,\ref{fig:Resultados-Generales} illustrates the fate of both planets calculated with our $N$-body code as a function of initial separations. Especially for separations close to the 2:1 and 3:2 mean motion resonances (MMRs; white dashed lines) the
  fate of the inner Neptune-mass planet is dramatically affected by the presence of the outer Jupiter. We find four cases in which both planets fall to the central star despite the Neptune-mass planet would have survived the RGB on its own (see the red-red squares at ($a_{\text{J}}$,$a_{\text{N}}$) =(2.5,1.95), (2.6, 1.85), (2.6, 1.90) and (2.6, 1.95)~au). In two simulations the
  Neptune-mass planet collides with the Jupiter-mass planet (yellow squares in Fig.\,\ref{fig:Resultados-Generales}) before the
  latter is engulfed by the central star. These collisions occurred in cases where both planets alone would not have survived. 
  We also find two cases where the Neptune-mass planet, which if it was on its own would have survived the RGB evolution of its
  host star, is ejected from the system due to close encounters with the outer Jupiter (small grey squares in
  Fig.\,\ref{fig:Resultados-Generales}). In one of these cases the Jupiter-mass planet survives while in the other one it is
  engulfed by the giant star. 

  The perhaps two most intriguing scenarios are, however, the following. On one hand, we find ``saviour cases" in which the
  Neptune-mass planet alone would not survive the RGB but is saved by its Jupiter-mass companion (two cases, small blue square
  above  big red square for red numbers on the y-axis in Fig.\,\ref{fig:Resultados-Generales}). On the other hand, we also find
  ``destroyer cases" where the Neptune-mass planet alone would not have been engulfed by the giant star but is killed by the outer
  Jupiter (six cases, small red square above  big blue square for blue numbers on the y-axis in Fig.\,\ref{fig:Resultados-Generales}).
  It is important to highlight that all these particular cases occur near the 3:2 and 2:1 MMR which cross the grid of the chosen
  semi-major axis. This does not imply that these cases are unlikely outliers. In contrast, Nature seems to have a preference
  for locating two consecutive planets close to the 3:2 and 2:1 MMR \citep[e.g.][]{Fabrycky2012,Trifonov2014}. That such a
  configuration can be destabilized during the RGB causing one of the planets to be ejected has previously been predicted by
  \citet{Voyatzis2013} who, however, did not take into account stellar tides. 
  Finally, we emphasize that due to the stochasticity of the passage throught the separatrix of the resonances, the results presented here can have different endings if the initial conditions are slightly changed, if a different timestep is considered, or even if a different computer is used \citep{Voyatzis2013, Folonier2014}.
 
\subsection{Destroyer and Savior scenarios}

\begin{figure*}
  \centering
    \includegraphics[width= 0.85\textwidth]{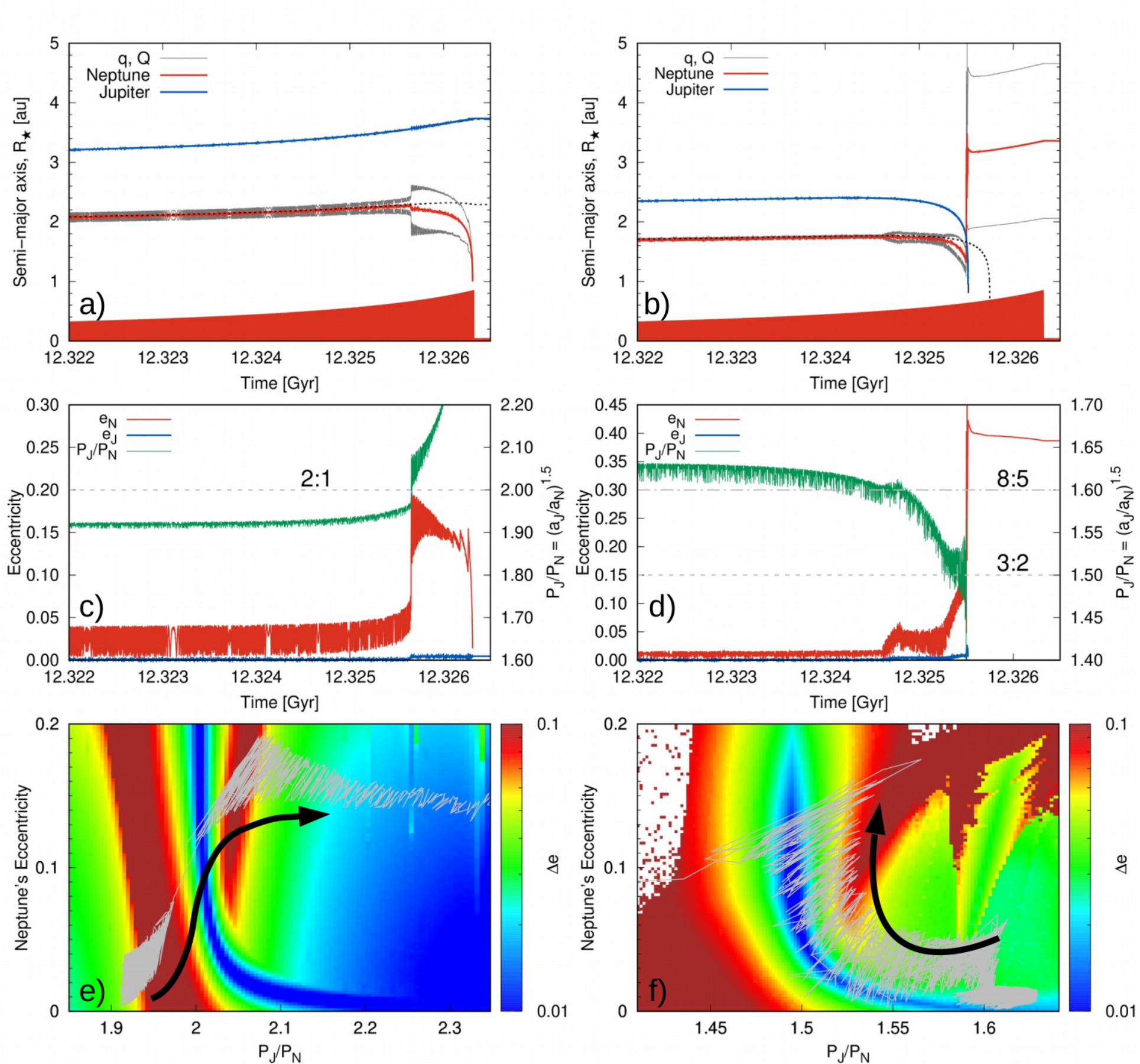}
    \caption{A detailed look at the ``destroyer" (left) and the ``saviour" (right) cases. The panels a and b show the time evolution of the Jupiter (blue) and the Neptune (red) semi-major axis for initial values of 3 and 1.95~au (left) and 2.2 and 1.6~au (right). The grey curves represent the Neptune's perihelion and aphelion, the short and long dashed lines the evolution if the planets were single. Panels c and d show the evolution of the eccentricities of both planets and their period ratio, with the location of the 2:1 MMR (left) and the 3:2 and 8:5 MMR (right) indicated by dotted lines. 
The bottom panels e and f show the corresponding dynamical maps. The colorscale represents the $\Delta e$ values in the ($P_{\text{J}}/P_{\text{N}},e$) plane for initial conditions in the vicinity of the 2:1 (left) and 3:2 MMR (right). 
Our N-body integrations are displayed by the grey lines. On the right (``saviour") the evolution is convergent (period ratio decreases) and $e$ increases as the system follows the 3:2 MMR. This leads to close encounters and finally the scattering event that saves the Neptune. In contrast, the evolution is divergent on the left and $e$ increases as the system crosses the 2:1 MMR. This causes the perihelion distance to decrease, tidal forces to increase, and finally the planet to fall into the star.    
}  
  \label{fig:KILLER-SURVIVAL}
\end{figure*}

The ``destroyer" and ``saviour" cases are fascinating scenarios that deserve a more detailed look. To that end we show the evolution of the separation and eccentricity for both cases as well as dynamical maps in Fig.\,\ref{fig:KILLER-SURVIVAL}. 
To construct the dynamical maps we set the Jupiters at their initial semi-major axis (3 and 2.2~au) and their eccentricity at its mean value attained in the first 100~Myr of the N-body integrations ($\sim5\times10^{-4}$). The ($a_{\text{N}},e_{\text{N}}$) plane was then divided in a $100\times100$ grid of initial conditions for the Neptune-mass planet. 
The semi-major axis ranged from 1.7 to 2\,au (left panel, ``destroyer" case) and 1.55 to 1.75\,au (right, ``saviour") and the eccentricity from 0 to 0.2. All these configurations were then integrated for 10,000 yr with the Neptunes represented by mass-less particles
and without considering stellar tides and stellar evolution. 

In the ``destroyer" case (panels a, c and e of Fig.\,\ref{fig:KILLER-SURVIVAL}) the surviving Jupiter causes the death of the Neptune-mass planet that would have survived the evolution of their host star on the RGB if it was alone (the simulation with a green circle in Fig.\,\ref{fig:Resultados-Generales}). 
Throughout the integration the planets do not experience close encounters. The evolution corresponds to a divergent migration as the period ratio increases. At first, the Neptune's eccentricity only slightly grows as their period ratio increases but instantly jumps to $\sim$0.15 as soon the planets cross the 2:1 resonance (see panels a and c of Fig.\,\ref{fig:KILLER-SURVIVAL}).
This increase in eccentricity causes the perihelion distance of the Neptune to significantly shrink. Stellar tides become more important at these smaller distances which further reduces decrease the separation until the planet finally falls into the envelope of the giant star. 

In the ``saviour" case the opposite occurs, i.e. the Neptune-mass planet survives although it would not if it was alone
(simulation with an orange circle in Fig.\,\ref{fig:Resultados-Generales}).
In this scenario, shown in detail in panels b, d and f of Fig.\,\ref{fig:KILLER-SURVIVAL}, 
the Jupiter-mass planet is initially set on a highway to hell and will indeed fall into the giant star, but not without saving its Neptune-mass companion. 

Despite their period ratio being initially far from the nominal value of the 3:2 MMR (i.e. $P_{\text{J}}/P_{\text{N}}\sim1.5$), 
the initial planet locations are within this resonance.
As a consequence, both planets evolve due to stellar tides following the apsidal corotation families, the bluish region in panel e where $\Delta_e\sim0$, \citep[see also][]{Giuppone+2013, Ramos+2015}. The migration in this case is convergent as the period ratio decreases. 
When the planets cross the 8:5 MMR, the Neptune's eccentricity is slightly enhanced but then remains constant for the next $\sim50,000$\,yr. Then, both planets continue evolving towards shorter separations (and period ratios) as stellar tides dominate over the effects of stellar mass loss until they get trapped in the 3:2 MMR. This trapping increases the Neptune's eccentricity to $\sim$0.15 which leads to close encounters between both planets. Just $\sim$0.8\,Myrs before the star reaches the tip of the RGB a planetary scattering event occurs during which the Neptune is kicked to an orbit with $a\sim$3.20~au and $e\sim$0.40. As the star is still losing mass and as at the increased orbital distance stellar tides are very inefficient, the Neptune's orbit expands further until the tip of the RGB reaching final orbital parameters of $a\sim$3.31~au and $e\sim$ 0.37. 

\subsection{Neptune RGB survivors}

\begin{figure}
  \centering
    \includegraphics[width= 0.47\textwidth]{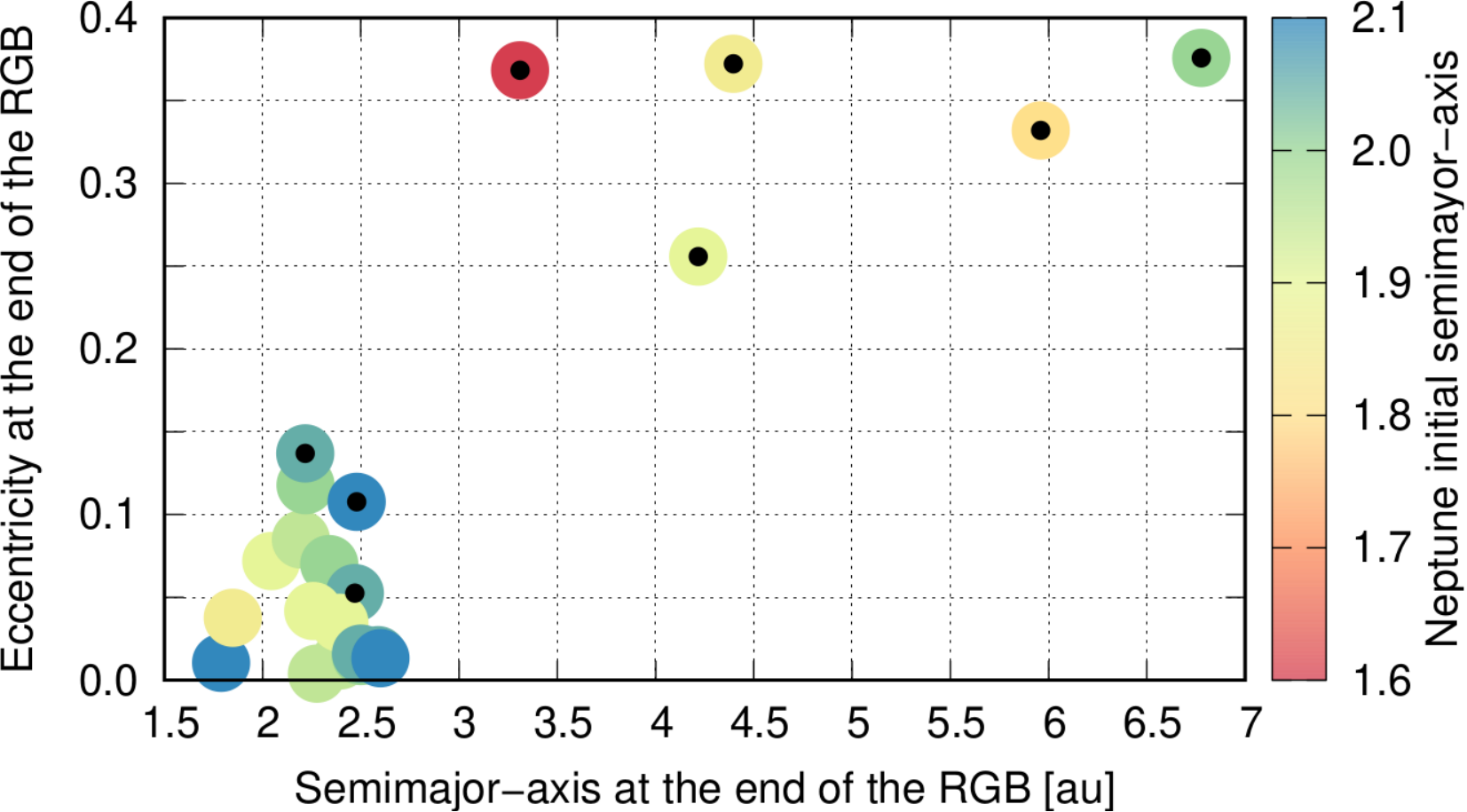}
    \caption{Semi-major axis vs eccentricity plane for the surviving Neptunes of Fig. \ref{fig:Resultados-Generales}. The Neptunes
      with black dots suffered close encounters with the Jupiter-mass planet.
  }
  \label{fig:Neptunes-Distribution}
\end{figure}

Inspecting the final orbital parameters of the surviving Neptune-mass planets from all the simulations shown in Fig.\,\ref{fig:Resultados-Generales}, we identify
two different populations as illustrated in Fig.\,\ref{fig:Neptunes-Distribution}. Surviving Neptunes with semi-major axis greater
than 3\,au and eccentricities larger than $\sim$0.25 result from planetary scattering events like the one described in the right panels of
Fig.\,\ref{fig:KILLER-SURVIVAL}, while those with lower eccentricities and semi-major axes result from gravitational interactions and
resonance crossings that enhanced their eccentricities but not enough to push them towards the star, as in the case of 
Fig.\,\ref{fig:KILLER-SURVIVAL} (left panels). 

Our findings therefore show that a significant fraction of planetary systems around white dwarfs might be shaped by gravitational
interactions, in particular resonances, occurring during the evolution of their host stars. Eccentric orbits of planets around
white dwarfs generated this way might play a significant role in scattering planetesimals or asteroids
\citep{Frewen2014,Smallwood2018, Antoniadou2019} and maybe even smaller planets closer to the white dwarf. It could therefore be
that the evolutionary scenarios discovered in this letter represent an important ingredient for understanding metal polluted white
dwarfs as well as the properties of planetary systems around white dwarfs. 

However, so far we have only considered two-planet systems while packed planetary systems with more planets might be a frequent
outcome of planet formation \citep{Guillon2017}. In fact, the most recent measurements indicate that on average planetary systems 
consist of more than three planets \citep{Zhu2018,Zink2019}.
In addition, our simulations only covered the first giant branch while the
progenitors of all currently known white dwarfs must have evolved as well through the AGB which would likely affect some of the
systems considered here \citep{MustillVillaver2012}. Finally, we only considered planets around a $1M_\odot$ star while most of the observed metal-polluted white dwarfs have more massive progenitors \citep{Koester2014} which suffer most of the radius expansion and mass-loss during AGB. We plan to overcome these limitations in future papers. 

\section{Conclusions}
  For the first time we combined stellar tides and multi-planet dynamics in an $N$-body code to study the evolution of a
  Neptune--Jupiter planetary system during the giant branch phases. We find that the fate of the Neptune-mass planet,
  located inside the Jupiter's orbit, can be significantly affected by the presence of the Jupiter-mass planet during and
  after the evolution of the host star on the RGB. When both planets are near a MMR, the eccentricity of the Neptune-mass
  planet is excited which affects its fate: Planets that would survive alone can be engulfed by the giant star and planets that
  would fall into the giant star if they were on their own can survive due to planet--planet scattering events. We also observe
  an increased eccentricity of Neptune-mass planets that survive the RGB evolution of their host star. While additional simulations
  covering different stellar and planetary masses and including AGB evolution are required, our results clearly show that
  gravitational interactions play an important role for fate of planets that are initially located at a few au from the star. In particular, resonances between planetary orbits occurring during the stars giant phases might be crucial to understand the architecture of planetary systems around white dwarfs. 

\acknowledgments
  We thank the anonymous referee for suggesting improvements to the
manuscript. MPR thanks Marcelo M.\ Miller Bertolami for useful discussions on stellar evolution aspects. MPR and MRS acknowledge
  support from FONDECYT (grants 3190336 and 1181404). MPR, MRS, JC and OMG are supported by the Iniciativa Cient\'{\i}fica
  Milenio (ICM) via the N\'ucleo Milenio de Formaci\'on Planetaria. MPR also thanks CONICYT project Basal AFB-170002. OMG is partially supported by PICT\,2016-0053 from ANPCyT, Argentina, and thanks IA-PUC for an invited research stay. DV gratefully acknowledges the support of the STFC via an Ernest Rutherford Fellowship (grant ST/P003850/1). CG acknowledges Mulatona Cluster from CCAD-UNC, which is part of SNCAD-MinCyT, Argentina.

%% To help institutions obtain information on the effectiveness of their 
%% telescopes the AAS Journals has created a group of keywords for telescope 
%% facilities.
%
%% Following the acknowledgments section, use the following syntax and the
%% \facility{} or \facilities{} macros to list the keywords of facilities used 
%% in the research for the paper.  Each keyword is check against the master 
%% list during copy editing.  Individual instruments can be provided in 
%% parentheses, after the keyword, but they are not verified.

%% Similar to \facility{}, there is the optional \software command to allow 
%% authors a place to specify which programs were used during the creation of 
%% the manuscript. Authors should list each code and include either a
%% citation or url to the code inside ()s when available.

%% Appendix material should be preceded with a single \appendix command.
%% There should be a \section command for each appendix. Mark appendix
%% subsections with the same markup you use in the main body of the paper.

%% Each Appendix (indicated with \section) will be lettered A, B, C, etc.
%% The equation counter will reset when it encounters the \appendix
%% command and will number appendix equations (A1), (A2), etc. The
%% Figure and Table counter will not reset.

\appendix

\section{Implementation of stellar tides}

  Different types of external forces affecting the evolution of a planet, like tides, interactions with a planetesimal disk,
 or disk torques, can be modeled by a Stokes non-conservative force as:
 \begin{eqnarray}
 \frac{{\text{d}}^2{\bf r}}{{\text{d}}t^2} = -C({\bf v} - \alpha {\bf v}_{\text{c}})
 \label{eq:}
\end{eqnarray}
\citep{Beauge2006}.
Here {\bf r} is the position vector referring to the star, {\bf v} is its velocity vector and {\bf v}$_{\text{c}}$ is the
circular velocity vector at the same point. $C$ and $\alpha$ are external coefficients. At first order in eccentricity
and for a single planet, the effects of the previous force in the semi-major axis and eccentricity of the body can be
described following \citet{Beauge1993}:
\begin{eqnarray}
 a(t) = a_0{\text{exp}}(-At), ~~~e(t) = e_0{\text{exp}}(-Et)
 \label{eq:exponentials}
\end{eqnarray}
where $a_0$ and $e_0$ are the conditions at the beginning of the integration and where $|A|$ and $|E|$ represent the
inverse of the e-folding times for $a$ and $e$, which can be computed as:
\begin{eqnarray}
 A = 2C(1-\alpha), ~~~E = C\alpha.
 \label{eq:variables}
\end{eqnarray}
To the first order, we can assume the right-hand sides of equations \ref{eq:semieje-tides} and \ref{eq:excentricidad-tides} 
as constant. Then, their solutions are formally given by Eq.\,\ref{eq:exponentials}, and $A=\left(\frac{\dot{a}}{a}\right)_{\text{t}}$
and $E=\left(\frac{\dot{e}}{e}\right)_{\text{t}}$ can be used to deduce the coefficients $C$ and $\alpha$ as:
\begin{eqnarray}
 C = \frac{1}{2}A + E, ~~~\alpha = \frac{E}{C}.
 \label{eq:}
\end{eqnarray}
Following this formalism the accelerations from tides were incorporated in our $N$-body code.

\begin{figure}
  \centering
    \includegraphics[width= 0.46\textwidth]{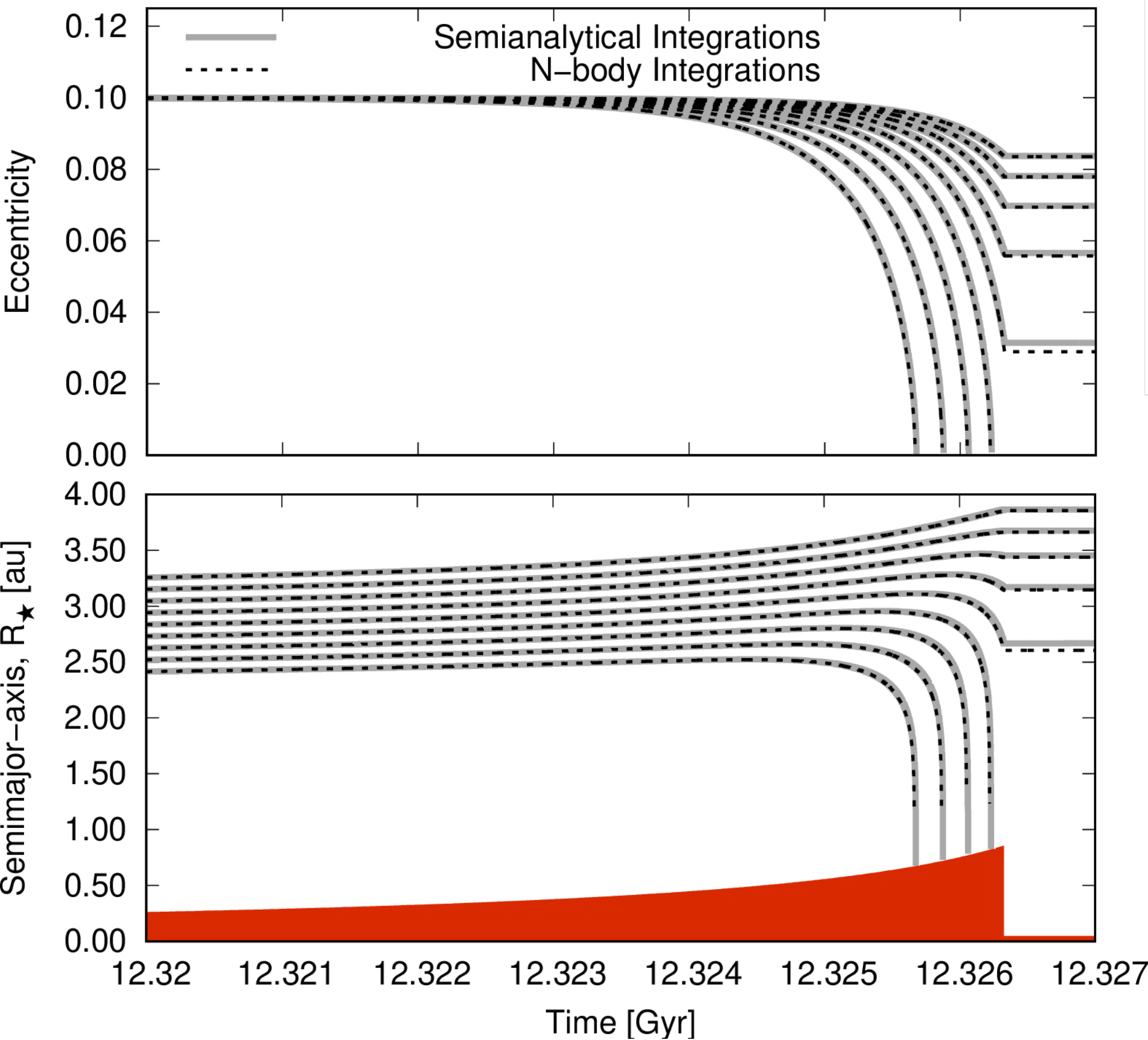}
    \vspace{-0.1cm}
    \caption{Comparison between evolution of the semi-major axis (bottom) and the eccentricity (top) of different orbits of a
      Jupiter-mass planet developed with the RKF integrator (grey solid line) and with the {\scriptsize{MERCURY}} $N$-body
      code (black dashed line). The red filled area represents the radius of the host star only for the last ∼6 Myr of its
      evolution towards the end of the RGB.}
  \label{fig:comparison}
  \vspace{-0.3cm}
\end{figure}

In order to test this implementation in {\scriptsize{MERCURY}} we evolved a single planet system until the central Solar-mass
star passed through the tip of the RGB, and compared the resulting orbital evolution with the RKF integrations. We use the same
evolutionary track as in Sec.\,\ref{sec:RKF-integrator}. The bottom panel of Fig.\,\ref{fig:comparison} shows results for a set
of simulations developed for a Jupiter-mass planet with different initial separations between 2.3~au and 3.1~au.  
It is clear that both kinds of integrations almost perfectly match. The eccentricity of these Jupiter-mass planets was
initially set to be 0.1 in order to also test the changes in this orbital parameter. The top panel of Fig. \ref{fig:comparison}
also shows a nearly perfect match in the evolution of the eccentricities.

\vspace{-0.3cm}

%% For this sample we use BibTeX plus aasjournals.bst to generate the
%% the bibliography. The sample63.bib file was populated from ADS. To
%% get the citations to show in the compiled file do the following:
%%
%% pdflatex sample63.tex
%% bibtext sample63
%% pdflatex sample63.tex
%% pdflatex sample63.tex

\bibliography{Biblio.bib}{}
\bibliographystyle{aasjournal}

%% This command is needed to show the entire author+affiliation list when
%% the collaboration and author truncation commands are used.  It has to
%% go at the end of the manuscript.
%\allauthors

%% Include this line if you are using the \added, \replaced, \deleted
%% commands to see a summary list of all changes at the end of the article.
%\listofchanges

\end{document}